\documentstyle[11pt]{article}
\begin{document}

 \def\beq{\begin{equation}}
 \def\eeq{\end{equation}}
 \def\bea{\begin{eqnarray}}
 \def\eea{\end{eqnarray}}
 \def\ul{\underline}
 \def\ni{\noindent}
 \def\nn{\nonumber}
 \def\wt{\widetilde}
 \def\wh{\widehat}
 \def\Tr{\mbox{Tr}\ }
 \def\tr{\,\mbox{tr}\,}                  
 \def\Tr{\,\mbox{Tr}\,}                  
 \def\Res{\,\mbox{Res}\,}                
 \renewcommand{\Re}{\,\mbox{Re}\,}       
 \renewcommand{\Im}{\,\mbox{Im}\,}       
 \def\lap{\Delta}                        

 \def\al{\alpha}
 \def\be{\beta}
 \def\ga{\gamma}
 \def\de{\delta}
 \def\ep{\varepsilon}
 \def\ze{\zeta}
 \def\io{\iota}
 \def\ka{\kappa}
 \def\la{\lambda}
 \def\na{\nabla}
 \def\ro{\varrho}
 \def\pa{\partial}
 \def\si{\sigma}
 \def\om{\omega}
 \def\ph{\varphi}
 \def\th{\theta}
 \def\te{\vartheta}
 \def\up{\upsilon}
 \def\Ga{\Gamma}
 \def\De{\Delta}
 \def\La{\Lambda}
 \def\Si{\Sigma}
 \def\Om{\Omega}
 \def\Te{\Theta}
 \def\Th{\Theta}
 \def\Up{\Upsilon}

 \vspace*{3mm}
 \begin{center}

 {\LARGE \sl Conformal symmetry, anomaly and effective action}
 \vskip 3mm

 {\LARGE \sl for metric-scalar gravity with torsion }

 \vskip 10mm

 {\bf J.A. Helayel-Neto}\footnote{ e-mail:
 helayel@cat.cbpf.br} \\ Centro Brasileiro de Pesquisas
 F\'{\i}sicas-CBPF-CNPq
 \\

 Universidade Cat\'olica de Petr\'opolis (UCP)
 \\
 \vskip 5mm
 {\bf A. Penna-Firme}\footnote{ e-mail:
 apenna@rio.nutecnet.com.br}
 \\
 Centro Brasileiro de Pesquisas F\'{\i}sicas-CBPF-CNPq \\
 Faculdade de Educacao da  Universidade Federal do
 Rio de Janeiro, (UFRJ)\\
 \vskip 5mm
 {\bf I. L. Shapiro}\footnote{ e-mail:
 shapiro@fisica.ufjf.br} \\ Departamento de Fisica -- ICE, Universidade
 Federal de Juiz de Fora\\
 Juiz de Fora, 36036-330, MG, Brazil
 \\
 Tomsk State Pedagogical University, Tomsk, Russia
 \end{center}
 \vskip 10mm

 
 {\centerline{\large \it {\sl ABSTRACT}}}
\vskip 3mm

\noindent
 We consider some aspects of conformal symmetry in a
 metric-scalar-torsion system. It is shown that, for some
 special choice of the action, torsion acts as a
 compensating field and the full theory is conformally
 equivalent to General Relativity on classical level. 
 Due to the introduction of
 torsion, this equivalence can be provided for the
 positively-defined gravitational and scalar actions.
 One-loop divergences arising from the scalar
 loop are calculated and both the consequent anomaly and
 the anomaly-induced effective action are derived.

 \vskip 10mm
 \section{Introduction}

 The studies in the framework of gravity with torsion have a long
 history and many interesting achievements.
 In particular, the relation between torsion
 and local conformal transformation
 (see, for example, \cite{histo1,hehl-PR} for a recent reviews
 with broad lists of references) is a topic of special relevance. 
 Torsion appears naturally in the effective action of string
 and, below the Planck scale, 
 should be treated, along with the metric, as part of the 
 gravitational background for the quantized matter fields.
 In this paper, we are going to investigate some details of conformal
 symmetry that come up in some metric-scalar-torsion
 systems. First, we briefly consider, following earlier
 papers \cite{deser,conf,duco}, a similar
 torsionless  system and then investigate the
 theory with torsion; proceeding further, we derive the one-loop
 divergences, conformal anomaly and the anomaly-induced
 effective action.

 Since we are going to consider the issues related to conformal
 transformation and conformal symmetry in gravity, it is
 worthy to make some general comments. The conformal transformations
 in gravity may be used for two different purposes.  First
 one can, by means of the
 conformal transformation, change the so-called conformal
 frame in the scalar-metric theory. Indeed, this choice is not
 arbitrary since the masses of the fields must be introduced.
 There may be various reasons to choose one or another frame as
 physical; one may find a detailed discussion on this issue in
 \cite{histo1}. All theories for which this consideration
 usually applies do not have conformal symmetry, even before the
 masses are introduced. Second, one can consider the
 theory with conformal symmetry.

 In the present paper, we are going to discuss
 only the second case, that is,
 massless theories with unbroken conformal symmetry.
 Then, the change of the conformal frame is nothing but the
 invariant (at least, at the classical level) choice of the
 dynamical variables. It is worth
 mentioning that the theories with conformal symmetry have
 prominent significance in many areas of modern theoretical
 physics. The most important, at least for the sake of
 applications, is that at the quantum level this symmetry
 is violated by anomaly. As examples of the application
 of the trace anomaly, one may recall the derivation of the
 effective
 equations for strings in background fields (which are nothing
 but conditions for the anomaly cancellation) \cite{CFMP},
 the study of the renormalization group flows (see, for example,
 \cite{anselmi} for the recent developments and list of references),
 the derivation
 of the black hole evaporation in the semiclassical approach
 \cite{black}, first inflationary model \cite{star} and
 some others (we recommend \cite{histo2} for a
 general review on conformal anomaly). Recently, the
 application of anomalies for $n=4$ has been improved
 by the use of the anomaly-induced effective action
 obtained long ago \cite{rei,frts}. In particular, this
 led to a better understanding of the anomaly-generated
 inflation \cite{anju} and allowed to perform a systematic
 classification
 of the vacuum states in the semiclassical approach to
 the black hole evaporation \cite{balsan}. Anomaly-induced
 effective action has been used to develop the quantum theory
 for the conformal factor \cite{antmot} and for the
 consequent study of the back reaction of gravity to the
 matter fields \cite{brv}.

 The generalization
 of the anomaly and the anomaly-induced action for the case
 of completely antisymmetric torsion has been given in
 \cite{buodsh}. Here we investigate a non-trivial
 conformal properties of the metric-scalar-torsion system,
 which modify the Noether identity corresponding to the
 conformal symmetry. This modification is due to the fact
 that, in the model under discussion, torsion
 {\sl does transform},
 while its antisymmetric part considered before \cite{buodsh}
 does not. As a result of this new feature, the conformal anomaly
 is not the anomaly of the Energy-Momentum Tensor trace,
 but rather the trace
 of some modified quantity. However, using a suitable
 parametrization for the fields one can, as it will be shown
 below, reduce the
 calculation of this new anomaly, and obtain the expressions
 for both the anomaly and the anomaly-induced effective action
 using corresponding results from \cite{rei,frts,buodsh}.

 For the sake of generality, we present
 all classical formulae in an $n$-dimensional space-time, for
 $n\neq 2$ (see the Appendix of \cite{duco} for the discussion
 of the special $n=2$-case in the torsion-free theory).
 The divergences and anomaly are all evaluated around $n=4$.

 \section{Brief review of the torsionless theory.}

 Our purpose here is to show that conformally invariant
 second-derivative $n$-dimensional metric-dilaton model
 is conformally equivalent to General Relativity. This
 equivalence has been originally demonstrated and discussed
 in \cite{deser} for the four-dimensional space and free scalar
 field, and later generalized in \cite{conf,duco} for the
 interacting scalar field in an arbitrary $n\neq 2$ dimensions.
 In this section, we review the result of \cite{deser,conf}
 presenting calculations in a slightly different manner.
 Our starting point will be the
 Einstein-Hilbert action with cosmological constant. 
 For further convenience we take this action with negative sign.
 \beq
 S_{EH}[g_{\mu\nu}] =
 \int d^nx \sqrt{- {\hat g}}\;
 \left\{ \, \frac{1}{G}\, {\hat R} + \La  \, \right\} \cdot
 \label{0.1}
 \eeq
 The above action depends on the metric $\,{\hat g}_{\mu\nu}\,$
 and we now set $\,{\hat g}_{\mu\nu} = g_{\mu\nu}\cdot e^{2\si(x)}$.
 In order to describe the local conformal transformation in
 the theory, one needs some relations between geometric
 quantities of the original and transformed metrics:
 \beq
 \sqrt{-{\hat g}} = \sqrt{-g}\;e^{n\si}\,, \;\;\;\;\;\;\;\;\;
 {\hat R} = e^{-2\si}\left[R -
 2(n-1)(\Box\si) -(n-2)(n-1)(\na\si)^2 \right] \cdot
 \label{n1}
 \eeq
 Substituting (\ref{n1}) into (\ref{0.1}), after
 integration by parts, we arrive at:
 \bea
 S_{EH}[g_{\mu\nu}] =
 \int d^nx \sqrt{-g}\,
 \left\{ \frac{(n-1)\,(n-2)}{G}\,e^{(n-2)\si}\,(\na\si)^2
 +\frac{e^{(n-2)\si}}{G}\,R + \La e^{n\si}\right \},
 \label{n2}
 \eea
 where $(\na \si)^2
 = g^{\mu\nu}\partial_\mu\si \partial_\nu\si $.
 If one denotes
 \beq
 \ph =
 e^{\frac{n-2}{2}\si}\,\cdot\,
 \sqrt{ \frac{8}{G}\cdot \frac{n-1}{n-2} } \,,
 \label{n3}
 \eeq
 the action (\ref{0.1}) becomes
 \beq
 S=\int d^nx \sqrt{-g} \left\{
 \frac{1}{2}\,(\na\ph)^2+\frac{n-2}{8(n-1)}R\ph^2+\La \left(
 \frac{G}{8}\cdot\frac{n-2}{n-1}\right)^{\frac{n}{n-2}}\cdot
 \ph^{\frac{2n}{n-2}}
 \right \} \cdot
 \label{n5}
 \eeq
 And so, the General
 Relativity with cosmological constant is equivalent to the
 metric-dilaton theory described by the action of eq. (\ref{n5}).
 One has to notice that the latter exhibits an extra local
 conformal symmetry, which compensates an extra (with respect to
 (\ref{0.1})) scalar degree of freedom. Moreover, (\ref{n5}) is a
 particular case of a family of
 similar actions, linked to each other
 by the reparametrization of the scalar or (and) the conformal
 transformation of the metric \cite{duco}. The symmetry
 transformation which leaves the action (\ref{n5}) stable,
 \beq
 g^{\prime}_{\mu\nu}=g_{\mu\nu}\cdot e^{2\rho(x)}
 \,,\, \, \, \, \, \, \, \, \, \, \,
 \ph^{\prime} = \ph \cdot e^{(1-\frac{n}{2})\rho(x)},
 \label{nnn}
 \eeq
 degenerates at $\,\,n \to 2\,\,$ and that is why this
 limit cannot be trivially achieved \cite{duco}.
 If we start from the positively defined gravitational action
 (\ref{0.1}), the sign of the scalar action (\ref{n5}) should
 be negative, indicating to the well known instability of the
 conformal mode of General Relativity (see \cite{wet,mod}
 for the recent account of this problem and further references).

 \section{Conformal invariance in metric-scalar-torsion theory}

 In this section, we are going to build up the conformally
 symmetric action with additional torsion field.
 Our notations are similar to the ones accepted in \cite{book}
 for the $n = 4$ case. Torsion is defined as
 $$
 {\wt \Gamma}^{\al}_{\be\ga} -  {\wt \Gamma}^{\al}_{\ga\be} =
 T^{\al}_{\be\ga}\,,
 $$
 where ${\wt \Gamma}^{\al}_{\be\ga}$ is the non-symmetric
 affine connection in the space provided by independent
 metric and torsion fields.
 The covariant derivative, $\,{\wt\na}_\al\,$, satisfies
 metricity condition,  $\,{\wt\na}_\al g_{\mu\nu}=0$.
 Torsion tensor may be decomposed into three irreducible parts:
 \beq
 {\wt \Gamma}_{\al , \be\ga} -  {\wt \Gamma}_{\al , \ga\be} =
 T_{\al , \be\ga} = \frac{1}{n-1} \left( T_{\beta}g_{\alpha\ga} -
 T_{\ga}g_{\alpha\beta} \right) - \frac{1}{3!(n-3)!}
 \varepsilon_{\alpha\beta\ga{\mu}_1 ... {\mu}_{n-3}}
 S^{{\mu}_1 ... {\mu}_{n-3}} + q_{\alpha\beta\la}\,.
 \label{irr}
 \eeq
 Here $\,T_\al = {T^\be}_{\al\be}\,$ is the
 trace of the torsion tensor $\,{T^\be}_{\al\ga}\,$.
 The tensor $\,S^{{\mu}_1 ... {\mu}_{n-3}}\,$
 is completely antisymmetric and, in the case of a
 purely antisymmetric torsion tensor, $\,T_{\al\be\ga}$ is
 its dual. $\,\varepsilon_{\alpha\beta\ga{\mu}_1 ... {\mu}_{n-3}}$
 is the maximal antisymmetric tensor density in the
 $\,n$-dimensional
 space-time. The sign of the $S$-dependent term corresponds
 to an even $\,n$. We notice that,
 in four dimensions, the axial tensor $\,S_{\mu_1 ... \mu_{n-3}}\,$
 reduces to the usual $S_\mu$ -- axial vector \cite{book}.
 In $n$ dimensions, the number of distinct components of the
 $\,S_{\mu_1 ... \mu_{n-3}}\,$ tensor is $\,\frac{n^3-3n^2+2n}{6}$.
 The tensor $\,{q^\be}_{\al\ga}\,$ satisfies, as in the
 $n=4$ case, the two constraints:
 $$
 \,{q^\be}_{\al\be}=0\,\,\,\,\,\,\,\,\,\,\,\,\, {\rm and }
 \,\,\,\,\,\,\,\,\,\,\,\,\,\,
 q_{\al\be\ga}\cdot \ep^{\al\be\ga\mu_1 ... \mu_{n-3}}=0
 $$
 and has $\,\frac{n^3-4n}{3}\,$ distinct components.
 We shall denote, as previously, Riemannian covariant derivative
 and  scalar curvature by $\na_\al$ and $R$ respectively,
 and keep the notation with tilde for the geometric quantities
 with torsion.

 The purpose of the present work is to
 describe and discuss conformal symmetry in the
 metric-scalar-torsion theory. It is well-known that
 torsion does not interact  minimally with scalar fields,
 but one can
 formulate such an interaction in a non-minimal way
 (see \cite{book} for the introduction). Moreover,
 this interaction between scalar field and torsion
 is necessary element of the renormalizable
 quantum field theory in curved space-time with torsion
 \cite{bush}. One may construct a general non-minimal action for the
 scalar field coupled to metric and torsion as below:
 \bea
 S =\int d^nx \sqrt{-g}\; \left\{
 \, \frac{1}{2} \;g^{\mu\nu}\partial_{\mu}\ph\;
 \partial_{\nu}\ph +  \frac{1}{2}\; \xi_i P_i \ph^2
 - \frac{\la}{4!}\,\ph^{\frac{2n}{n-2}} \, \right\}\,.
 \label{n7}
 \eea
 Here, the non-minimal sector is described by five structures:
 \bea
 P_1 = R \,,\,\,\,\,
 P_2=\na_\al T^\al\,,\,\,\,\,
 P_3=T_\al\;T^\al \,,\,\,\,\,
 P_4=S_{{\mu}_1 ... {\mu}_{n-3}}\;S^{{\mu}_1 ... {\mu}_{n-3}}
 \,,\,\,\,\,\,
 P_5=q_{\al\be\ga}\,q^{\al\be\ga}\,;
 \label{nn8}
 \eea
 in the torsionless case the only $\,\xi_1 R\,$ term is present.
 $\xi_i$ are the non-minimal parameters which
 are typical for the theory in external field. Renormalization
 of these parameters is necessary to remove corresponding
 divergences
 which really take place in the interacting theory \cite{bush}.
 Furthermore, the renormalizable theory always includes some
 vacuum action. This action must incorporate all structures
 that may show up in the counterterms. The general expression
 for the vacuum action in the case of a metric-torsion background
 has been obtained in \cite{chris}; it contains 168 terms.
 In fact, one can always reduce this number, because not all the
 terms with the allowed dimension really appear as counterterms.
 The discussion of the vacuum renormalization
 for the external gravitational field with torsion has been
 previously given in \cite{buodsh} for a purely antisymmetric
 torsion. In the present
 article, we will be interested in the special case of the (\ref{n7})
 action, which possesses an interesting conformal symmetry.
 The appropriate expression for the corresponding vacuum action
 will be given in Section 4, after we derive vacuum counterterms.

 Using the decomposition (\ref{irr}) given above,
 the equations of motion for the torsion tensor can
 be split into three independent equations
 written for the components
 $T_\al,\,S_\al,\,q_{\al\be\ga}$; they yield:
 \beq
 T_\al=\frac{\xi_2}{\xi_3}\,\cdot\, \frac{\na_\al\ph}{\ph}
 \,,\,\,\,\,\,\,\,\,\,\,\,\,
 \,S_{\mu_1 ... \mu_{n-3}} = q_{\al\be\ga} = 0\,.
 \label{n9}
 \eeq
 Replacing these expressions back into the action (\ref{n7}),
 we obtain the on-shell action
 \bea
 S = \int d^nx \sqrt{-g}\; \left\{
 \frac{1}{2} (1-\frac{\xi^2_2}{\xi_3})\;
 g^{\mu\nu}\partial_{\mu}\ph \;
 \partial_{\nu}\ph +  \frac{1}{2}\; \xi_1 \ph^2R
 -\frac{\la}{4!}\ph^{\frac{2n}{n-2}} \right \}\,,
 \label{onshell}
 \eea
 that can be immediately reduced to (\ref{n5}), by an
 obvious change of variables, whenever
 \beq
 \xi_1=\frac{1}{4}\left(1-\frac{\xi^2_2}{\xi_3}\right)\frac{n-2}{n-1}\,.
 \label{n8}
 \eeq
 Therefore, we notice that the version of the
 Brans-Dicke theory with torsion (\ref{n7})
 is conformally equivalent to General Relativity (\ref{0.1}) provided
 that the new condition (\ref{n8}) is satisfied and no sources for
 the components $T_\al,\,S_{\mu_1 ... \mu_{n-3}}\,$ and
 $\,q_{\al\be\ga}$  of the torsion tensor are included.
 In fact, the introduction of
 external conformally covariant sources for
 $\,S_{\mu_1 ... \mu_{n-3}}$ , $\,{q^{\alpha}}_{\beta\gamma}\,$
 or to the transverse component of $\,T_{\al}\,$
 does not spoil the conformal symmetry.

 One has to remark, that the theory with torsion provides, for
 $\,\frac{\xi^2_2}{\xi_3}-1 > 0$, the equivalence of the
 positively defined scalar action (\ref{n7}) to the
 action (\ref{0.1}) with the negative sign. The negative
 sign in (\ref{0.1}) signifies, in turn, the positively defined
 gravitational action. Without torsion one can achieve positivity
 in the gravitational action only by the expense of the negative
 kinetic energy for the scalar action in (\ref{n5}).
 Thus, the introduction of torsion may lead to some
 theoretical advantage.

 The equation of
 motion for $T_\al$ may be regarded as a constraint that fixes
 the conformal transformation for this vector to be consistent
 with the one
 for the metric and scalar. Then, instead of (\ref{nnn}), one has
 \beq
 g^{\prime}_{\mu\nu}=g_{\mu\nu}\cdot e^{2\rho(x)}\,,\, \, \, \, \,
 \, \, \, \, \, \,
 \ph^{\prime} = \ph \cdot e^{(1-\frac{n}{2})\rho(x)}
 \,,\, \, \, \, \, \, \, \, \, \, \,
 T^{\prime}_\al =
 T_\al + (1-\frac{n}{2})\cdot\frac{\xi_2}{\xi_3}\cdot\partial_\al\rho(x)
 \label{mmm}
 \eeq
 Now, in order to be sure about the number of degrees of freedom in
 this theory,
 let us compute the remaining field equations for the whole set of
 fields. From this instant we consider free theory and put the
 coupling constant $\,\la=0$, because scalar self-interaction
 does not lead to the change of the qualitative results.

 The dynamical equations for the theory encompass eqs. (\ref{n9})
 along with
 \bea
 &&\frac{1}{2}\xi_1\ph^2\left(R_{\al\be}-\frac{1}{2}g_{\al\be}R \right)
 -\frac{1}{4}g_{\al\be}(\na\ph)^2 +
 \frac{1}{2}\na_\al\ph\na_\be\ph \nonumber \\ &+&
 \frac{1}{2}\xi_2\ph^2 \left(T_\al T_\be - \frac{1}{2}g_{\al\be}T_\rho
 T^\rho
  \right) + \frac{1}{2}\xi_3\ph^2\left( \na_\al T_\be -\frac{1}{2}
  g_{\al\be}\na_{\la}T^\la \right)=0
 \label{teq}
 \eea
 and
 \bea
 \left(\Box-\xi_1R-\xi_2\na_\la T^\la -\xi_3T_\la T^\la\right) \ph=0\,.
 \eea
 In case of the on-shell torsion (\ref{n9}), equations
 (\ref{teq}) reduce to:
 \bea
 & &\frac{1}{8}\frac{n-2}{n-1}\phi^2
 \left(R_{\mu\nu}-\frac{1}{2}g_{\mu\nu}R
 \right)-\frac{1}{4}g_{\mu\nu}(\na\phi)^2+
 \frac{1}{2}\na_\mu\phi\na_\nu\phi=0\,, \\
 & & \Box \phi-\frac{1}{4}\left(\frac{n-2}{n-1}\right)\,R\,\phi=0\,.
 \label{equa}
 \eea
 Taking the trace of the first equation, it can be readily
 noticed that
 this equation is exactly the same as the one for scalar field.
 This  indeed justifies our
 procedure for replacing (\ref{n9}) into the action.
 It is easy to check, by direct inspection,
 that even off-shell, the theory with
 torsion (\ref{n7}), satisfying the relation (\ref{n8}), may be
 conformally invariant whenever we define the transformation
 law for the torsion trace according to (\ref{mmm}):
 and also postulate that the other pieces of the torsion,
 $S_\mu$ and ${q^\al}_{\be\ga}$, do not transform.
 The quantities $\,\sqrt{-g}\,$ and $\,R\,$ transform as in
 (\ref{n1}).
 One may introduce into the action other conformal invariant
 terms depending on the torsion. For instance:
 \beq
 S = - \frac{1}{4}\,
 \int d^nx\,\sqrt{-g}\,\,\ph^{\frac{2\cdot(n-4)}{n-2}}\,
 T_{\al\be}\, T^{\al\be},
 \eeq
 where $T_{\al\be}=\partial_\al T_\be-\partial_\be T_\al$.
 This term reduces to the usual vector action when $n\to 4$.
 It is not difficult to propose other conformally
 invariant terms containing
 other components of the torsion tensor in (\ref{irr}).

 One can better understand the equivalence between GR and conformal
 metric-scalar-torsion theory (\ref{n7}), (\ref{n8})
 after presenting an alternative form for the symmetric action.
 It proves a useful way to divide the torsion trace
 $T_{\mu}$ into longitudinal and transverse parts:
 \beq
 T_{\mu} = T^{\bot}_{\mu} + \frac{\xi_2}{\xi_3}\,\pa_\mu\,T \,,
 \label{dividi}
 \eeq
 where $\,\na^\mu T^{\bot}_{\mu} = 0\,$ and $\,T\,$ is the
 scalar component of the torsion trace, $T_\mu$. The
 coefficient $\,\frac{\xi_2}{\xi_3}\,$ has been introduced for
 the sake of convenience. Under the conformal transformation
 (\ref{mmm}), the transverse part is inert and the
 scalar component transforms as $\,T^\prime = T - \si$. $\,$
 Now, we can see that the conformal invariant components,
 $\,S,T^{\bot}\,$ and $\,q$, appear in the action (\ref{n7})
 in the combination
 \beq
 {\cal M}^2 =  \xi_3\,(T^{\bot}_{\mu})^2
 + \xi_4\, S_{\mu_1\mu_2 ... \mu_{n-3}}S^{\mu_1\mu_2 ... \mu_{n-3}}
 + \xi_5\, q_{\mu\nu\la}^2\,,
 \label{combi}
 \eeq
 which has the conformal transformation,
 $\,{{\cal M}^2}^\prime
 \rightarrow e^{(2-n)\rho (x)}\cdot {\cal M}^2\,$
 similar to that of the square of the scalar field.
 In fact, the active compensating role of the torsion trace,
 $T_{\mu}$, is accommodated in the scalar mode $T$.
 Other components of
 the torsion tensor, including $T^{\bot}_{\mu}$, appear only
 in the
 form (\ref{combi}) and allow to create some kind of
 {\it conformally covariant} "mass". On the mass-shell, this
 "mass" disappears
 because all its constituents,
 $\,T^{\bot}_{\mu}, S_{\mu_1 ...\mu_{n-3}}\,$ and
 $\,q_{\mu\nu\la}$, vanish.

 The next observation is that all torsion-dependent terms may be
 unified in the expression
 $$
 P = -\frac{n-2}{4(n-1)}\,\frac{\xi^2_2}{\xi_3}\,R +
 \xi_2\,(\na_\mu T^\mu) +
 \xi_3\,T_\mu T^\mu +
 \xi_4\, S_{\mu_1\mu_2 ... \mu_{n-3}}^2
 + \xi_5\,q_{\mu\nu\la}^2
 $$
 \beq
 = - \,\frac{n-2}{4(n-1)}\,\frac{\xi_2^2}{\xi_3}\,
 \left[ R - \frac{4(n-1)}{n-2}\,(\na T)^2
 - \frac{4(n-1)}{n-2}\,\Box T\right] + {\cal M}^2 \,.
 \label{P}
 \eeq
 It is easy to check that the transformation law for this
 $\,P\,$ is the same as the one for $\,{\cal M}^2$.
 Using new definitions, the invariant action becomes
 \beq
 S_{inv} = \int d^nx \sqrt{-g}\; \left\{
 \,\frac{1}{2} \;g^{\mu\nu}\partial_{\mu}\ph\;
 \partial_{\nu}\ph +  \frac{n-2}{8(n-1)}\;R\ph^2
 + \frac12\,P\,\ph^2 \right\}\,.
 \label{inv}
 \eeq
Let us make a change of variables in the last action:
$$
T = \ln \psi\,,
$$
with obvious transformation law, $\psi' = \psi\cdot e^{-\sigma}$,
for the new scalar $\,\psi$.
After some small algebra, one can cast the action (\ref{inv})
in the
form of a 2-component conformally invariant sigma-model:
\beq
S_{inv} = \frac12\, \int d^nx \sqrt{-g}\; \left\{
\,- \,\ph \De_2 \ph \,+\,{\cal M}^2\ph^2
\,+\, \frac{\xi_2^2}{\xi_3}\,(\ph^2\psi^{-1})\De_2 \psi
\,\right\}\,,
\label{inv-new}
\eeq
where
$$
\De_2 = {\Box} - \frac{n-2}{4(n-1)}\,R
$$
is a second-derivative conformally covariant operator acting on
scalars. It is easy to check that if one takes, for
example, the conformally flat metric,
$g_{\mu\nu}=\eta_{\mu\nu}\cdot \chi^2(x)$, one arrives
at three-scalar non-linear sigma-model, but two
of these three scalars are dependent.
The last expression for the action (\ref{inv-new})
shows explicitly the conformal invariance of the
action and also confirms the conformal covariance of the quantity
$\,P$.

One may construct a trivial 2-scalar sigma-model
conformally equivalent to GR, by making a substitution
$\,\si \rightarrow \si_1 + \si_2\,$ in the action (\ref{n2}),
and after regarding $\,\si_1(x)\,$ and $\,\si_2(x)\,$
as distinct fields.
A detailed analysis shows that the theory (\ref{inv-new})
can not be reduced to such a 2-scalar sigma-model
by a change of variables. In order to understand why this
is so, we need one more representation for
the metric-scalar-torsion action with
local conformal symmetry.

Let us  start, once again, from the action (\ref{n7}), (\ref{n8})
and perform only part of the transformations (\ref{mmm}): \beq \ph
\rightarrow  \ph^{\prime} = \ph \cdot e^{(1-\frac{n}{2})\rho(x)}
\,,\, \, \, \, \, \, \, \, \, \, \, T_\al \rightarrow
T^{\prime}_\al = T_\al +
(1-\frac{n}{2})\cdot\frac{\xi_2}{\xi_3}\cdot\partial_\al\rho(x)\,.
\label{partofmmm} \eeq Of course, if we supplement
(\ref{partofmmm}) by the transformation of the metric, we arrive
at (\ref{mmm}) and the action does not change. On the other hand,
the results of Section 2 suggest that (\ref{partofmmm}) may lead
to an alternative conformally equivalent description of the
theory. Taking $\,\ph \cdot e^{(1-\frac{n}{2})\rho(x)} =
\frac{8(n-1)}{G\,(n-2)}\,\left(1-\frac{\xi^2_2}{\xi_3}\right)
 = const$,
we obtain, after some algebra, the following action:
$$
 S = \frac{1}{G}\,\int d^nx \sqrt{-g}\;R +
$$
\beq
+ \frac{4(n-1)}{G(n-2)(1-{\xi^2_2}/{\xi_3})}\,
\int d^nx \sqrt{-g}\;\left\{
\xi_4 S_{\mu_1 ... \mu_{n-3}}^2
+ \xi_5 q_{\mu\nu\la}^2
+ \xi_3 \left(T_\al-\frac{\xi_2}{\xi_3}\,\na_\al\ln\ph
\right)^2\right\}\,.
\label{oneq}
\eeq
This form of the action does not contain interaction between
the curvature and the
scalar field. At the same time, the latter field
is present until we use the equations of motion
(\ref{n9}) for torsion.
Torsion trace looks here like a Lagrange multiplier, and only
using its corresponding equation of motion (and also for other
components of torsion), we can obtain the action of GR.
It is clear that one can arrive at the same action
(\ref{oneq}) making the transformation of the metric
as in (\ref{mmm}) instead of (\ref{partofmmm}).

 To complete this part of our consideration, we mention that
 the direct generalization
 of the Einstein-Cartan theory including an extra scalar may be
 conformally equivalent to General Relativity, provided
 that the non-minimal parameter takes an appropriate
 value. To see this, one uses the relation
 \beq
 {\wt R}= R - 2\na_\al \, T^\al - \frac{n}{n-1}
 T_\al\,T^\al + \frac{1}{2}\,
 q_{\alpha \beta \gamma} q^{\alpha \beta \gamma} +
 \frac{1}{4}\, \cdot \, \frac{1}{3!(n-3)!}\, S^{\al_1 ... \al_{n-3}}\,
 S_{\al_1 ... \al_{n-3}}
 \label{curvator}
 \eeq
 and replace it into the "minimal" action
 \beq
 S_{ECBD} =
 \int d^nx \sqrt{-g}\, \left\{ \frac{1}{2}\,g^{\mu\nu}\partial_\mu\ph\,
 \partial_\nu\ph+\frac{1}{2}\,\xi\,{\wt R}\ph^2 \right\} \,.
 \label{mini}
 \eeq
 It is easy to see that the condition (\ref{n8}) is satisfied
 for the special value
 \beq
 \xi =  \frac{n(n-2)}{8(n-1)}\,.
 \label{special}
 \eeq
 In particular, in the four-dimensional case, the symmetric
 version of the theory corresponds to $\xi = \frac13$,
 contrary to the famous $\xi = \frac16$ in the
 torsionless case. The effect of changing conformal value of
 $\xi$ due to the non-trivial transformation of torsion
 has been discussed earlier in \cite{payo} (see also further
 references there).

In the next section we shall see how the non-trivial
conformal transformation for torsion changes the Noether
identity and the quantum conformal anomaly.

\section{ Divergences, anomaly and induced effective action}

In four dimensions, by integrating over the free scalar field
(even without self-interaction), one meets vacuum
divergences and the resulting trace anomaly breaks conformal
invariance. As it was already mentioned in the Introduction,
the anomaly and its application is one of the most important
aspects of the conformal theories. The anomaly
is a consequence of the quantization procedure, and it
appears due to the lack of a completely invariant regularization.
In particular, trace anomaly is usually related to the one-loop
divergences \cite{duff}. At the same time, one has to be
careful, because the non-critical use of this relation may,
in principle, lead to mistakes.

Consider the renormalization and anomaly for the
conformal metric-scalar theory with torsion formulated above.
The renormalizability of the theory requires the vacuum action
to be
introduced, which has to be (as it was already noticed in
Section 3) of the form of the possible
counterterms. The total four-dimensional action including the
vacuum term can be presented as:
\beq
S_t = S_{inv} + S_{vac}\,,
\label{tot}
\eeq
where $S_{inv}$ has been given in (\ref{inv}) and the form of the
vacuum action will be established later.

Before going on to calculate the divergences and anomaly, one has
to write a functional form for the conformal symmetry. It is easy
to see that the Noether identity corresponding to (\ref{mmm})
looks like 
\beq 
\,2 g_{\mu\nu}\,\frac{\de
S_t}{\de g_{\mu\nu}} +\frac{\xi_2}{\xi_3}\,\pa_\mu\,\frac{\de
S_t}{\de T_\mu} - \ph\,\frac{\de S_t}{\de \ph}  \,=\,0 \,.
\label{neuth} 
\eeq 
Now, if we are considering both metric and
torsion as external fields, and only the scalar as a quantum
field, in the vacuum sector we meet the first two terms of
(\ref{neuth}). This means (due to the invariance of the vacuum
divergences) that the vacuum action may be chosen in such a way
that 
\beq 
-\sqrt{-g}\,{\cal T}\, = 
\,2 g_{\mu\nu}\,\frac{\de S_{vac}}{\de g_{\mu\nu}}
+\frac{\xi_2}{\xi_3}\,\pa_\mu\,\frac{\de S_{vac}}{\de T_\mu}
  \,=\,0 \,.
\label{vacu}
\eeq
The new form (\ref{vacu}) of the conformal Noether
identity indicates to a serious modification in the
conformal anomaly. In the theory under discussion, the
anomaly would mean $\,<{\cal T}>\,\neq\,0$ instead
of usual $\,<{T}^\mu_\mu>\,\neq\,0$. Therefore, at
first sight, we meet here some special case and one can not
directly use the relation between the one-loop
counterterms and the conformal anomaly derived in \cite{duff},
because this relation does not
take into account the non-trivial transformation law
for the torsion field. It is reasonable to remind that,
for the case of completely antisymmetric torsion
discussed in \cite{buodsh}, this problem did not show up
just because $S_\mu$ is inert under conformal
transformation.

Let us formulate a more general statement
about conformal transformation and anomaly.
When the classical metric background is
enriched by the fields which do not transform, the
Noether identity corresponding to the conformal symmetry
remains the same as for the pure metric background.
In this case
one can safely use the standard relations \cite{duff}
between divergences and conformal anomaly.
However, if the new background field has a non-trivial
transformation law, one has to care about possible
modifications of the Noether identity and consequent
change of anomaly.

In our case, the anomaly is modified, because
it has a new functional form, $\,<{\cal T}>\,\neq\,0$.
One can derive this new anomaly using, for instance, the
methods described in \cite{duff} or \cite{book}.
However, it is possible to
find $\,<{\cal T}>\,$ in a more economic way, using
some special decomposition of the background fields.
As we shall see, the practical calculation of a new
anomaly and even the anomaly-induced action may be
reduced to the results known from \cite{duff,rei,frts}
and especially \cite{buodsh}, where the theory of the
antisymmetric torsion was investigated.

Our purpose is to change the background variables in such
a way that the transformation of torsion is absorbed by
that of the metric. The crucial
observation is that $P$, from (\ref{P}),
transforms\footnote{As a
consequence, the action $\,\int\sqrt{-g}P\phi^2\,$ is conformal
invariant. This fact has been originally discovered in
\cite{obukhov}.} under (\ref{mmm}) as $P^\prime = P \cdot
e^{-2\rho(x)}$. Therefore, the non-trivial transformation of
torsion is completely absorbed by $P$. Since $P$ only
depends on
the background fields, we can present it in any
useful form. One can imagine, for instance, $P$
to be of the form
$\,P=g^{\mu\nu}\,\Pi_\mu\,\Pi_\nu\,$ where vector
$\,\Pi_\mu\,$
doesn't transform, and then the calculation readily reduces
to the case of an antisymmetric torsion \cite{buodsh}.
In particular, we can now use standard results for
the relation between divergences and anomaly \cite{duff}.

In the framework of
the Schwinger-DeWitt technique, we find the
1-loop counterterms in the form
 \bea
 \Ga^{(1)}_{div} = - \frac{\mu^{n-4}}{(4\pi)^2\,(n-4)}\,
 \int d^nx\sqrt{-g}\,\left\{\frac{1}{120}\,C^2-\frac{1}{360}\,E
 + \frac{1}{180}\,{\Box}R + \frac{1}{6}\,{\Box} P
 + \frac{1}{2}\,P^2 \right \}\,,
 \label{div}
 \eea
Here 
$$
C^2 = C_{\mu\nu\al\be}C^{\mu\nu\al\be}\,\,\,\,\,\,\,\,\,\,\,\,\,\,\,\,\,\,
{\rm and} \,\,\,\,\,\,\,\,\,\,\,\,\,\,\,\,\,\,
E = R_{\mu\nu\al\be}R^{\mu\nu\al\be} - 4R_{\mu\nu\al\be}R^{\mu\nu}
+R^2
$$ 
are the square of the Weyl tensor, and  
the scalar integrand of the Gauss-Bonnet term. The eq.
(\ref{div}) gives, as a by-product, the list of necessary terms in
the vacuum action. Taking into account the arguments presented
above, we can immediately cast the anomaly under the form
 \bea
 <{\cal T}> \,=\, - \frac{1}{(4\pi)^2}\,\left[
 \frac{1}{120}\,C_{\mu\nu\al\be}\,C^{\mu\nu\al\be}-\frac{1}{360}\,E
 + \frac{1}{180}\,{\Box}R + \frac{1}{6}\,{\Box} P
 + \frac{1}{2}\,P^2 \right]\,.
 \label{tracean}
 \eea
One may proceed and, following \cite{rei,frts,buodsh}, derive the
conformal non-invariant part of the effective action of the
vacuum, which is responsible for the anomaly (\ref{tracean}).
Taking into account our previous
treatment of the conformal transformation of torsion, we consider
it is hidden inside the quantity $P$ of eq. (\ref{P}),
and again imagine $P$ to be of the form
$\,P=g^{\mu\nu}\,\Pi_\mu\,\Pi_\nu\,$. Then
the equation for the effective action
$\Ga [g_{\mu\nu},\Pi_\al]\,$ is
\footnote{We remark that this equation is valid only for
the "artificial" effective action
$\Ga [g_{\mu\nu},\Pi_\al]\,$, while the effective
action in original variables $g_{\mu\nu}, T^\al_{\be\ga}$
would satisfy the modified equation (\ref{vacu}). The
standard form of the equation for the effective action
is achieved through
the special decomposition of the external fields.}
\beq
- \frac{2}{\sqrt{-g}}\,g_{\mu\nu} \frac{\de\,\Ga}{\de
g_{\mu\nu}} {} = <{\cal T}>\,.
\label{mainequation}
\eeq
In order
to find the solution for $\,\Ga$, we can factor out the conformal
piece of the metric ${g}_{\mu\nu} = {\bar g}_{\mu\nu}\cdot
e^{2\si}$, where ${\bar g}_{\mu\nu}$ has fixed determinant
and put $\,P = {\bar P}\cdot e^{-2\si(x)}$, that
corresponds to $\,{\bar \Pi}_\al = \Pi_\al$.
This transformation for artificial variable $\Pi_\al$
is identical to the one for the $S_\al$ pseudovector, that
is the case considered in \cite{buodsh}.
Then the result can be obtained directly from the
effective action derived in \cite{buodsh}, and we get
$$
{\Ga}\, = \,S_c[{\bar g}_{\mu\nu}; {\bar P}]
\,- \,\frac{1}{12}\cdot\frac{1}{270 (4\pi)^2}\,\int d^4 x
\sqrt{-g (x)}\,R^2(x) + \frac{1}{(4\pi)^2}\,
\int d^4 x\sqrt{-{\bar g}}\,\left\{
\,\si\cdot [ \frac{1}{120}\,{\bar C}^2 -
\right.
$$
\beq
\left.
- \frac{1}{360}\,({\bar E} - \frac23\,{\bar \na}^2 {\bar R})
+ \frac12\,{\bar P}^2 ] +
 \frac{1}{180}\,\si {\bar \De}\si
-\frac16\,({\bar \na}_\mu\si)\,{\bar \na}^\mu {\bar P}
+\frac16\,{\bar P}({\bar \na}_\mu\si)^2\,\right\}\,,
\label{efac}
\eeq
where $S_c[{\bar g}_{\mu\nu}; {\bar P}]\,$ is an unknown
functional of the metric  ${\bar g}_{\mu\nu}(x)$ and $\,{\bar P}\,$,
which acts as an integration constant for any solution of
(\ref{mainequation}).

Now, one has to rewrite (\ref{efac}) in
terms of the original field variables, $g_{\mu\nu}, {T^\al}_{\be\ga}$.
Here, we meet a small problem, because we only have, for the moment,
the definition $\Pi_\al = {\bar \Pi}_\al$ for the artificial variable
$\Pi_\al$, but not for the torsion. Using the previous result
(\ref{mmm}), we can define
\beq
 {T^\al}_{\be\ga} = {{\bar T}^\al}_{\,\,\be\ga}
-\frac{1}{3}\cdot\left[\,\de^\al_\ga\,\pa_\be\si -
\de^\al_\be\,\pa_\ga\si\,\right]\,, \label{mmm-new} \eeq so that
$\,{{\bar T}^\al}_{\,\,\be\ga}\,$ is an arbitrary tensor. Also, we
call ${\bar T}^\al = {\bar g}^{\al\be}\,{\bar T}_\be$ etc. Now, we
can rewrite (\ref{efac}) in terms of metric and torsion components
$$ {\Ga}\, = \,S_c[{\bar g}_{\mu\nu}; {{\bar T}^\al}_{\,\,\be\ga}]
\,- \,\frac{1}{12}\cdot\frac{1}{270 (4\pi)^2}\,\int d^4 x \sqrt{-g
(x)}\,R^2(x) + $$$$ + \frac{1}{(4\pi)^2}\, \int d^4 x\sqrt{-{\bar
g}}\,\left\{ \,+ \frac{1}{180}\,\si {\bar \De}\si +
\frac{1}{120}\,{\bar C}^2\,\si - \frac{1}{360}\,({\bar E} -
\frac23\,{\bar \na}^2 {\bar R})\,\si \right. $$$$ \left. +
\frac{1}{72}\,\si\,\left[ \,-\frac{\xi^2_2}{\xi_3}\,{\bar R} +
6\xi_2\,({\bar \na}_\mu {\bar T}^\mu) + 6\xi_3\,{\bar T}_\mu {\bar
T}^\mu  + 6\xi_4\, {\bar S}_{\mu}{\bar S}^{\mu} + 6\xi_5\,{\bar
q}_{\mu\nu\la}{\bar q}^{\mu\nu\la} \right]^2 + \right. $$ \beq
\left. \frac16\,\left[({\bar \na}^2\si\,+\, ({\bar
\na}_\mu\si)^2\,\right]\cdot \left[
\,-\frac{\xi^2_2}{\xi_3}\,{\bar R} + 6\xi_2\,({\bar \na}_\mu {\bar
T}^\mu) + 6\xi_3\,{\bar T}_\mu {\bar T}^\mu + 6\xi_4\, {\bar
S}_{\mu}{\bar S}^{\mu} + 6\xi_5\,{\bar q}_{\mu\nu\la}{\bar
q}^{\mu\nu\la} \right]\,\right\}\,, \label{efac-final} \eeq This
effective action is nothing but the generalization of the similar
expressions of \cite{rei,frts,buodsh} for the case of general
metric-torsion background and conformal symmetry described in
Section 3. The curvature dependence in the last two terms appears
due to the non-trivial transformation law for torsion.

\section{Conclusions.}

We have considered the conformal properties
of the second-derivative metric-torsion-dilaton gravity, on both
classical and quantum level.
The main results, achieved above, can be summarized as follows:
\vskip 2mm

i) If the parameters $\,\xi_i\,$ of the general metric-torsion-dilaton
theory (\ref{n7}) satisfy the relation
(\ref{n8}), there is conformal invariance of the new
type, more general than the one considered in \cite{obukhov,payo}
and different from the ones considered in \cite{bush,book}.
\vskip 2mm

ii) In this case the vector trace of torsion plays the role of a
compensating field, providing the classical
conformal on-shell equivalence between this
theory and General Relativity. The conformal invariant
scalar-metric-torsion action can be presented in
alternative forms (\ref{inv}), (\ref{inv-new}), (\ref{oneq}),
while (\ref{inv-new}) is some new nonlinear $n$-dimensional
conformal invariant sigma-model. If the nonminmal
parameters satisfy an additional relation
$\,\,{\xi^2_2} > {\xi_3}$, the equivalence holds between
the positively defined General Relativity and scalar field
theory. This indicates to the stabilization of the
conformal mode in gravity with torsion and may indicate
that the introduction of torsion can be some
useful alternative to
other approaches to this problem (see \cite{wet,mod}).
\vskip 2mm

iii) The conformal Noether identity (\ref{neuth})
indicates that the conformal anomaly with torsion
is not the anomaly of the trace of the Energy-Momentum
Tensor, but rather the anomaly of some modified quantity
$\,{\cal T}$. But, using the appropriate decomposition of
the background variables, the calculation of this new
anomaly can
be readily reduced to the use of the known results of \cite{duff}
and \cite{buodsh}, and the derivation of the
anomaly-induced effective action can be done through
the similar decomposition of variables and the use
of the known results of \cite{rei,frts,buodsh}.
As in other known cases, the induced action (\ref{efac-final})
breaks the conformal equivalence between metric-torsion-scalar
action and (\ref{0.1}) and therefore may lead to the
nontrivial applications to inflationary cosmology.

\vskip 6mm
 \noindent
{\bf Acknowledgments.}
I.Sh. is grateful to the CNPq for permanent support.
His work was partially supported by RFFI
(project 99-02-16617). A.P.F. and J.A.H.-N.
are grateful to G. de Berredo
Peixoto for discussions and careful reading of the work.


\end{document}